\documentstyle[epsfig]{elsart}
\def\bk{{\bf k}}
\def\be{\begin{equation}}
\def\ee{\end{equation}}
\def\bea{\begin{eqnarray}}
\def\eea{\end{eqnarray}}
\def\l({\left(}
\def\r){\right)}

\def\bk{{\bf{k}}}

\def\dst{\displaystyle{\phantom{|}}}
\def\ov{\over\dst}

\begin{document}
\begin{frontmatter}

\title{Coulomb wave function corrections for \\
$n$-particle Bose-Einstein correlations }

\author[Mainz]{E. O.\ Alt\thanksref{erwin},} 
\author[KFKI]{T. Cs{\"o}rg\H o\thanksref{tamas},}
\author[Lund]{B. L{\"o}rstad\thanksref{bengt},}
\author[Lund]{J. Schmidt-S{\o}rensen\thanksref{janus}}
\address[Mainz]{Institut f\"ur Physik, Universit\"at Mainz, D-55099 Mainz, Germany}
\address[KFKI]{MTA KFKI RMKI, H-1525 Budapest 114, POB 49, Hungary}
\address[Lund]{Physics Department, Lund University, S-221 00 Lund, POB 118, Sweden}

\thanks[erwin]{Email: Erwin.Alt@uni-mainz.de}
\thanks[tamas]{Email: csorgo@sunserv.kfki.hu}
\thanks[bengt]{Email: bengt@quark.lu.se}
\thanks[janus]{Email: janus@quark.lu.se}

\date{\today}

\begin{abstract}
The effect of multi-particle Coulomb 
final state interactions on higher-order intensity correlations
is determined in general, based on
a scattering wave function which is solution of the $n$-body 
Coulomb Schr\"odinger equation in (a large part of) the asymptotic 
region of $n$-body configuration space. 
In particular, we study Coulomb effects on the $n$-particle 
Bose-Einstein correlation functions of similarly charged
particles  and remove  a systematic error as big as 100 \% 
from higher-order multi-particle Bose-Einstein correlation
functions.
\end{abstract}
\begin{keyword}
n-body Coulomb, Bose-Einstein correlations, heavy ion, elementary particle
\end{keyword}
\end{frontmatter}

\section{Introduction}

One of the most fundamental quests of high-energy physics is
the determination of the phase diagram of strongly interacting
matter. At high densities and/or temperatures the
quarks are expected to be liberated from their confinement within
hadrons and a new phase of matter, the Quark Gluon Plasma (QGP), to be
formed.  In order to explore this new phase of matter,
the Relativistic Heavy Ion Collider (RHIC) has been constructed
at Brookhaven National Laboratory, to collide Au + Au nuclei 
at $\sqrt{s} = 200$ AGeV center of mass energy.

One of the new features of RHIC physics will be the production
of 600 - 1200 charged pions per unit rapidity. Due to this reason,
the PHENIX and STAR detectors will be able to determine not only
single particle spectra and the two-particle Bose-Einstein
correlations, but also the higher-order Bose-Einstein correlation
functions, which turn out to be essential in order to distinguish
fully chaotic and partially coherent particle sources from each
other~\cite{pnhalo}. As partial coherence is a fundamental aspect
of quantum fields,  and it can be related to a possible Bose-Einstein
condensation of pion wave-packets to the wave-packet with the smallest
energy in the rest frame of the source~\cite{cstjz}, the determination
of the higher-order Bose-Einstein correlation functions at RHIC
is of great theoretical 
interest~\cite{pnhalo,Cramer,nhalo,biyajima,suzuki,heinz,axel,zhang}, 
as well as a great experimental challenge. 

However,  Coulomb (and possibly strong) final state
interactions of pions play an important r{\^o}le in shaping the final
multi-particle Bose-Einstein correlation functions. 
As no consistent and systematic treatment of 
the final state interaction of a charged multi-boson
system is available in the literature, the experimental
removal of the Coulomb effects from the $n$-particle Bose-Einstein
correlation functions is based hitherto only on some ad-hoc generalization
of the Gamow formula to the multi-particle case.

In the present Letter we propose a straight-forward method for
a {\it systematic} quantum-mechanical treatment of 
Coulomb final state interactions in higher-order Bose-Einstein 
correlation functions. Although the validity of the method, 
described below, is limited to a certain, albeit large kinematic
domain $(\Omega_{0}^{(n)})$, due to the fact that the exact solution of even
the 3-body Coulomb scattering problem is beyond presently available means,
we think that the results presented here represent a first important 
step towards establishing 
a link between few-body physics and Bose-Einstein correlations
 in high-energy multi-particle physics. Especially,
the new Coulomb wave-function corrections indicate that 
the generalized Gamow correction method would make 
a factor of two error in the 5-th order Bose-Einstein correlation
functions, if the radius parameters were in the 5 - 10 fm range,
as characteristic for heavy ion collisions.
If the characteristic radius parameters were as small as 1 fm,
 the characteristic size in reactions of high-energy particle physics,
the generalized Gamow factors would be acceptable,
at a 10 \% level of precision in the 5-th order Bose-Einstein
correlation functions.

The strength of higher-order correlation functions
increases much slower for partially coherent particle sources
than for incoherent sources with an unresolvable
halo of long-lived resonances~\cite{Cramer,pnhalo}.
Even if the strength of the second- and third-order
Bose-Einstein correlations were similar in a
partially coherent and another incoherent particle
source, the strength of the 5-th order correlation
functions would be  a factor of 2 different  between
the partially coherent and the fully chaotic cases~\cite{pnhalo}.
In order to distinguish these scenarios, the 
Coulomb final state interactions must be corrected for,
and the error on the Coulomb correction must be kept
under control.

 In a recent Letter, we have presented a
 refined treatment of the 3-body Coulomb correction problem~\cite{acls},
 with application to new high-energy heavy-ion data 
 by the NA44 experiment~\cite{janus-na44}.
 In this work, we generalize this 3-body Coulomb wave function
 integration method to the case 
 of $n$-particle Coulomb corrections. 

\section{Bose-Einstein $n$-particle correlations and final state interactions} 

Let us summarize some basic properties of the Bose-Einstein
$n$-particle correlation functions using only the generic aspects of 
their derivation, and establish a link between the theory of
final state interactions in few body physics and the theory
of Bose-Einstein correlations in high-energy particle and nuclear physics.

The $n$-particle Bose-Einstein correlation function is defined as
\be 
	C_n(\bk_1, \cdots , \bk_n ) = {\dst N_n(\bk_1, \cdots , \bk_n) \ov
		N_1(\bk_1) \cdots N_1(\bk_n) },
\ee
where $N_n(\bk_1, \cdots , \bk_n) $ is the $n$-particle inclusive invariant
momentum distribution, while $N_1(\bk_1)$ is the single-particle invariant
momentum distribution.
It is quite remarkable that this complicated object, that carries quantum
mechanical information on the phase-space distribution of particle 
production as well as on possible partial coherence of the source,
can be expressed in a relatively simple, straight-forward manner
both in the analytically solvable pion-laser model of Refs.~\cite{pratt,cstjz} 
as well as in the generic boosted-current formalism of 
Gyulassy and Padula~\cite{gyu-pa} as
\be
	C_n(\bk_1, \cdots , \bk_n ) = {\dst
		\sum_{\sigma^{(n)}} \prod_{i = 1}^n G(\bk_i,\bk_{\sigma_i})	
		\ov \prod_{i=1}^n G(\bk_i,\bk_i) }, \label{e:m2}
\ee
where $\sigma^{(n)}$ stands for the set of permutations of indices
$(1, 2, \cdots, n)$ and $\sigma_i $ denotes the element replacing
element $i$ in a given permutation from the set of $\sigma^{(n)}$,
and, regardless of the details of the two different derivations,
\be
 G(\bk_i,\bk_j) = \langle a^{\dagger}(\bk_i) a(\bk_j) \rangle
\ee
stands for the expectation value of $a^{\dagger}(\bk_i) a(\bk_j) $.
In the boosted-current formalism, the derivation is based on the
assumptions that (i) the bosons are emitted from a semi-classical source,
where currents are strong enough so that the recoils due to radiation
can be neglected, (ii) the particle sources are 
an incoherent random ensemble of such currents, described by
a boost-invariant formulation~\cite{gyu-pa}, and (iii) that 
the particles propagate as free plane waves after production. 
However, a formally similar result is obtained when particle production
happens in a correlated manner, and even final state interactions
between the produced particles are allowed for,
generalizing the results of Refs.~\cite{cstjz,jzcst,brood}.

In the pion-laser model, the 
$n$-particle exclusive invariant momentum distributions read as
\be
	N_n^{(n)} 
	( \bk_1, \cdots , \bk_n )  =  
	\sum_{\sigma^{(n)}} \prod_{i = 1}^n G_1(\bk_i, \bk_{\sigma_i})
\ee
with
\be
	 G_1(\bk_i\bk_j)  = \mbox{Tr} \{\hat \rho_1 a^{\dagger}(\bk_i) a(\bk_j)\}, 
		\label{e:bas1}
\ee
where $\hat \rho_1$ is the single-particle density matrix in the limit
when higher-order Bose-Einstein correlations are negligible.
One can show~\cite{jzcst,zhang}, that the $n$-particle
inclusive spectrum has a similar structure,
if the multiplicity distribution is Poissonian in the rare gas limit: 
\bea
	N_n
	( \bk_1, \cdots , \bk_n ) & = & 
	\sum_{\sigma^{(n)}} \prod_{i = 1}^n G(\bk_i,\bk_{\sigma_i})\\
	 G(\bk_i,\bk_j) & = & \sum_{n=1}^\infty G_n(\bk_i,\bk_j) \label{e:bas}.
\eea
The functions $G_n(\bk_i,\bk_j)$ can be considered as representatives of 
order $n$ symmetrization effects in exclusive events, 
see refs.~\cite{pratt,cstjz,jzcst} for  more detailed definitions. 
The function $G(\bk_i,\bk_j)$ can be considered as the expectation value of
$a^{\dagger}(\bk_i) a(\bk_j) $ in an inclusive sample of events,
and this building block includes all the higher order symmetrization effects. 
In the relativistic Wigner-function formalism, 
in the plane wave approximation $G(\bk_1,\bk_2)$ 
can be rewritten as
\bea
	G(\bk_1,\bk_2) & = & \int d^4 x \,S(x,K_{12})\, \exp(i q_{12}\cdot x)
		\label{e:gwig}\\
	K_{12} & = & 0.5 (k_1 + k_2) \\
	q_{12} & = & k_1 - k_2,
\eea
	where a four-vector notation is introduced,
	$k = (\sqrt{m^2 + \bk^2}, \bk)$, and $a \cdot b$ stands for the 
	inner product of four-vectors.
	Due to the mass-shell constraints, i.e.\ $E_\bk = \sqrt{m^2 + \bk^2}$,
	$G$ depends only on 6 independent momentum components.
	In any given frame, the boost-invariant decomposition
	of Eq.~(\ref{e:gwig}) can be rewritten into the following,
	seemingly not invariant form:
\bea
	G(\bk_1,\bk_2) & = & \int d^3 {\bf x}\,\, S_{{\bf K}_{12}}({\bf x})\,\,
		\exp(i {\bf q}_{12} {\bf x}),
		\label{e:gwignr}\\
	S_{{\bf K}_{12}}({\bf x}) 
	& = & \int dt\, \exp( i  {\mbox {\boldmath $\beta$}}_{K_{12}} 
	{\bf q}_{12} t)\,\, S( {\bf x}, t, K_{12}), \\
	{\mbox {\boldmath $\beta$}}_{K_{12}} & = & ({\bf k}_1 + \bk_2) /(E_1 + E_2).
\eea
	Note that  the relative source function
	$S_{{\bf K}_{12}}({\bf x})$ 
	reduces to  a simple time integral over the source function
	$S(x,K)$ in the frame where the 	mean momentum of the pair
	(hence the pair velocity ${\mbox {\boldmath $\beta$}}_{K_{12}}$) vanishes.

	If $n$ particles are emitted with similar momenta,
	so that their $n$-particle Bose-Einstein correlation
	functions may be non-trivial, Eqs.~(\ref{e:bas1},\ref{e:bas})
	will form the basis for evaluation of 
	the Coulomb  and strong final state interaction effects 
	on the observables.
	On this level, all the correlations are build up from correlations
	of pairs of particles. This is due to the specific form of
	the density matrix that includes just the right amount of stimulated
	emission to make a further calculation straight-forward.
	Note also that a similar result can be obtained in 
	the semi-classical boosted-current formalism, where the 
	particle production has negligible effect on the elementary
	source of pion production (bremsstrahlung-like radiation).

	Let us point out that the exact solution of multi-particle 
	Bose-Einstein symmetrization in the pion laser model resulted
	in a Poisson cluster picture~\cite{jzcst,cstjz}. This implies
	that in the rare gas limit, without Coulomb or other final state
	interactions, the multi-boson correlations appear only 
	as random admixture of a small amount of correlated pairs to 
	independently distributed single particles.
	As the density increases, also the fraction of correlated pairs increases
	and the admixture of independently distributed clusters of particle
	triplets, quartets, and higher-order $n$-tuples becomes correspondingly 
	more important. The result of refs.~\cite{jzcst,cstjz} 	indicates
	that below the onset of Bose-Einstein condensation, 
	a fully symmetrized multi-boson system can be considered 
	as a convolution of independently distributed clusters 
	of particle $n$-tuples, and it is natural to 
	apply Coulomb corrections within such clusters
	of particles only. We shall also discuss that, when one of the 
	particles becomes separated from its cluster, the relevant
	$n$-particle Coulomb correction factor will reduce to
	the Coulomb correction factor of a smaller cluster that contains  
	the remaining $n-1$ particles. 
 
\section{Quantum mechanical treatment of the Coulomb $n$-body problem} 

In order to treat correctly the Coulomb corrections to the $n$-particle 
correlation function, knowledge of the $n$-body Coulomb scattering wave 
function is required. 
We restrict ourselves to the case that the transverse momenta of all
the particles in the final state in their center of mass are small enough 
to make a nonrelativistic approach sensible. 
Hence the problem consists in finding 
the solution of the $n$-charged particle Schr\"odinger 
equation when all $n$ particles are in the continuum. 

Consider $n$ distinguishable 
particles with masses $m_{i}$ and charges 
$e_{i},\,i = 1,2,$ $\cdots $, $n$. Let ${\rm {\bf {x}}}_{i}$ and 
${\rm {\bf {k}}}_{i}$ denote the coordinate and momentum 
(three-)vectors, respectively, of particle $i$. From these 
we construct in the usual manner the relative coordinate 
${\rm {\bf {r}}}_{ij} = 
{\rm {\bf {x}}}_{i} - {\rm {\bf {x}}}_{j}$ and the relative 
momentum ${\rm {\bf {k}}}_{ij} = 
(m_j {\rm {\bf {k}}}_{i} - m_i {\rm {\bf {k}}}_{j})/{(m_{i} +m_{j})}$ 
between particles $i$ and $j$, the corresponding reduced 
mass being $\mu_{ij} = m_{i}m_{j}/(m_{i} +m_{j})$.

The $n$-particle Schr\"odinger equation reads
\begin{eqnarray}
\left\{ H_0 + \sum_{i < j=1}^{n} V_{ij} - E \right\} 
\Psi^{(+)}_{ {\rm {\bf {k}}}_{1} \cdots {\rm {\bf {k}}}_{n} }
({\rm {\bf {x}}}_{1},\cdots, {\rm {\bf {x}}}_{n}) = 0, 
\label{nse}
\end{eqnarray}
where
\begin{eqnarray}
E = \sum_{i=1}^{n}\frac{{\rm {\bf {k}}}_{i}^2}{2m_{i}} > 0
\end{eqnarray}
is the total kinetic energy for $n$ particles in the continuum. 
$H_0$ is the free Hamilton operator and
\begin{eqnarray}
V_{ij}({\rm {\bf {r}}}_{ij}) = V_{ij}^S({\rm {\bf {r}}}_{ij}) 
+ V_{ij}^C({\rm {\bf {r}}}_{ij}) 
\end{eqnarray}
the interaction potential between particles $i$ and $j$, consisting of 
a strong but short-range ($V_{ij}^S$) plus the 
long-range Coulomb interaction ($V_{ij}^C({\rm {\bf {r}}}_{ij}) 
= e_i e_j/\mid {\rm {\bf {r}}}_{ij} \mid$). Equation (\ref{nse}) 
has to be complemented by 
the complete set of boundary conditions in order that a unique 
solution be obtained.

Already for $n= 3$, 
the exact numerical solution of the Schr\"odinger equation (\ref{nse}) 
for $E>0$ is beyond present means, partly for principal and partly 
for practical reasons. For a brief discussion of the related difficulties 
see \cite{a98b}. But, at least the complete set of boundary conditions 
to be imposed is nowadays known analytically \cite{r72,am92}, in the form 
of the explicit solutions of the Schr\"odinger equation in all asymptotic 
regions of the three-particle configuration space. Apart from the trivial 
two-cluster region relevant for an asymptotic configuration containing 
only two particles one of which is a bound state of two particles, the 
asymptotic solution takes its simplest form in the asymptotic region 
conventionally denoted by $\Omega _{0}$ and characterized by the fact 
that - roughly speaking - all three interparticle distances 
become uniformly large, i.e.\ all $\mid {\rm {\bf {r}}}_{ij} 
\mid \to \infty $ (for 
a precise definition of the various asymptotic regions see \cite{am92}). 
There exist three more asymptotic regions $\Omega _{ij},\,i<j=1,2,3,$ 
which are pertinent to situations characterized by final state 
interactions between particles $i$ and $j$. But the appropriate 
asymptotic solutions are rather more complicated.  From the physical 
point of view the union of all these regions $\Omega _{0} \bigcup 
\Omega _{12} \bigcup \Omega _{13} \bigcup \Omega _{23}$ is relevant 
for the complete breakup into three free particles. As has 
been shown in \cite{acls}, in spite of the lack of
an exact solution of the three-body Schr\"odinger equation 
in the whole three-body configuration space, already 
knowledge of the asymptotic solution in $\Omega _{0}$ led 
to a systematic, well controlled extraction of Coulomb effects 
in the three-particle Bose-Einstein correlation measurements,
in contrast to earlier, ad-hoc 3-body Coulomb correction methods.

In the final states of heavy-ion
reactions, where a large number of charged particle tracks appear, 
the mutual, macroscopically large separation  of tracks
is one of the criteria of a clean measurement. This suggests 
that in order to study $n$-body correlation functions, again 
knowledge of the wave function in $\Omega_{0}^{(n)}$, the region 
in $n$-particle configuration 
space where all interparticle distances become uniformly large, i.e.,
$\mid {\rm {\bf {r}}}_{ij} \mid 
\to \infty$ for all values of $(ij)$, may be sufficient. 
Here, uniform divergence of interparticle distances in $\Omega_{0}^{(n)}$ 
means roughly that $0 < \mid {\rm {\bf {r}}}_{ij} \mid / 
\mid {\rm {\bf {r}}}_{kl} \mid < \infty$ for asymptotically 
large times for any two arbitrarily chosen particle pairs, 
although the interparticle distances themselves diverge 
for any pair. One immediate consequence of this definition is that in 
$\Omega_{0}^{(n)}$ the short-range interaction parts $V_{ij}^S$ 
play no r{\^o}le any longer and can thus be neglected.

For want of an exact $n$-particle Coulomb scattering wave function an approximate solution of Eq.\ (\ref{nse}) is sought. For this purpose,
let us introduce the continuum solution of the two-body 
Coulomb Schr\"odinger equation as
\medskip
\begin{eqnarray} 
	\left\{ - \frac{\Delta_{{\rm {\bf{r}}}_{ij}}}
	{2 \mu_{ij}} + V_{ij}^C({\rm {\bf{r}}}_{ij}) - \frac {{\rm {\bf {k}}}_{ij}^{2}}
	{2 \mu_{ij}} \right\} 
	{\psi}_{ {\rm{\bf{k}}}_{ij}}^{C(+)}({\rm {\bf{r}}}_{ij}) = 0, 
	\label{2cse} 
\end{eqnarray}
describing the relative motion of the two particles $i$ and $j$ with 
energy ${\rm {\bf {k}}}_{ij}^{2}/2 \mu_{ij}$. The explicit solution is 
\medskip
\begin{eqnarray} 
	{\psi}_{ {\rm {\bf {k}}}_{ij}}^{C(+)}
	({\rm {\bf {r}}}_{ij}) 
	&=& N _{ij} \; e^{ i {\rm {\bf {k}}}_{ij} 
	{\rm {\bf {r}}}_{ij}} 
	F[- i \eta _{ij}, 1; i ( \mid {\rm {\bf {k}}}_{ij} \mid 
\mid {\rm {\bf {r}}}_{ij} \mid - {\rm {\bf {k}}}_{ij} 
	{\rm {\bf {r}}}_{ij})], \label{2pwfn}
\end{eqnarray} 
with $N_{ij} = e^{- \pi \eta_{ij}/2}\, 
\Gamma (1 + i \eta_{ij}), $ and $\eta_{ij}= 
e_{i} e_{j}\mu_{ij}/ \mid {\rm {\bf {r}}}_{ij} \mid $ being the appropriate 
Coulomb parameter. $F[a,b;x]$ is the confluent hypergeometric function and 
$\Gamma (x)$ the Gamma function. 
With it the following ansatz for an approximate $n$-particle Coulomb 
wave function is made 
\begin{eqnarray} 
\Psi^{(+)}_{ {\rm {\bf {k}}}_{1}, \cdots ,
 {\rm {\bf {k}}}_{n} }({\rm {\bf {x}}}_{1}, \cdots ,{\rm {\bf {x}}}_{n}) \; 
\sim \; \prod_{i < j=1}^n 
{\psi}_{ {\rm {\bf {k}}}_{ij}}^{C(+)}({\rm {\bf {r}}}_{ij}).
\label{aswn}
\end{eqnarray} 

This ansatz can be justified by the following arguments. \\
(i) The wave function (\ref{aswn}) is asymptotically correct in 
the asymptotic region $\Omega_0^{(n)}$; that is, it is the leading 
term if all interparticle separations go to infinity, of the (unknown) 
exact solution of the Schr\"odinger equation (\ref{nse}) \cite{r72}. 
Of course, for non-asymptotic particle separations it represents a 
theoretically not compulsory though plausible extrapolation. \\
(ii) In the formal, time-dependent scattering theory the basic 
object is the M{\O}LLER operator which maps the free $n$-particle 
state onto the corresponding scattering state. The mathematically 
rigorous definition of the $n$-charged particle M{\O}LLER 
operator \cite{dollard64} requires, in contrast to the case 
of purely short-range interactions between the particles, the 
introduction of a `renormalization factor'. The latter has 
the form of a product of $n(n-1)/2$ renormalization factors each of which is
 appropriate for the definition of the M{\O}LLER operator for 
one of the possible pairings of the charged particles. Obviously, 
the ansatz (\ref{aswn}) of the $n$-particle wave function is 
consistent with this renormalization prescription. \\
(iii) By suitably decomposing the (stationary) M{\O}LLER operator 
of an $n$-body system into a chain of M{\O}LLER operators of 
subsystems with fewer interacting particles, a wave function 
of the type (\ref{aswn}) has been suggested as a lowest-order term
of an $n$-particle Coulomb wave function in \cite{Briggs90} 
to be used in all of configuration space. The 
assumptions entering were neglect of genuine higher-than-two 
particle correlations in the wave function, which is justified 
in $\Omega_0^{(n)}$, and restriction of all two-particle scatterings 
onto their respective energy shells. \\
(iv) The wave function (\ref{aswn}) coincides for any selected 
particle triplet with the form pertinent to the given pre-selected triplet, 
if the corresponding interparticle distances diverge \cite{acls}. \\
(v) For $n=3$, such an approximate wave function has been proposed in 
Refs.~\cite{m77,bbk89}. Although it ceases to be solution 
of the Schr\"odinger equation (\ref{nse}) for non-asymptotic 
values of the relative coordinates, it is nevertheless 
widely used, with considerable success, to calculate cross 
sections for the ionisation of hydrogen atoms by (energetic) 
electron impact.

The foregoing discussion makes it clear that the ansatz 
(\ref{aswn}) is justifiable only for sufficiently large 
interparticle separations, a condition which is not easily 
translated into an experimentally accessible criterion. However, 
for $n=3$ such a criterion has been established, namely that 
the total kinetic energy $E_{total}^{(3)}$ of three particles 
of unit charge be at least $0.2 \hbar c/R_{G}(\mbox{fm})$ which 
equals $10$ MeV for a typical source size $R_{G} = 4$ fm \cite{acls}. 
Thus, assuming also for an arbitrary number $n$ of particles 
(again for simplicity 
taken to have unit charge) the total kinetic energy $E_{total}^{(n)}$ 
to be equally distributed over the relative kinetic energies 
between each pair, the latter criterion generalises to the 
condition $E_{total}^{(n)} \geq 0.033 n(n-1) \hbar c/R_{G}(\mbox{fm})$ 
MeV. Hence, for the following we {\it assume} the $n$-particle Coulomb wave 
function to be given everywhere as
\begin{eqnarray}
\Psi^{(+)}_{ {\rm {\bf {k}}}_{1} \cdots {\rm {\bf {k}}}_{n} }
({\rm {\bf {x}}}_{1}, \cdots ,{\rm {\bf {x}}}_{n}) \; &\approx& \; 
\tilde \Psi^{(+)}_{ {\rm {\bf {k}}}_{1} \cdots {\rm {\bf {k}}}_{n} }
({\rm {\bf {x}}}_{1}, \cdots ,{\rm {\bf {x}}}_{n}) \nonumber \\
\; & := &\; \sqrt{{\cal N}^{(n)}} \; \prod_{i < j=1}^n 
{\psi}_{ {\rm {\bf {k}}}_{ij}}^{C(+)}({\rm {\bf {r}}}_{ij}), \; 
\label{aswn1} \\
\null &\null &
 \mbox{for}\, E_{total}^{(n)} \geq 0.066 \,\frac{n(n-1)}{2} 
\frac{\hbar c}{R_{G}\mbox{(fm)}} \;\mbox{MeV},
\end{eqnarray}
where ${\cal N}^{(n)}$ is an undetermined overall normalization constant. 
This is the building block for a properly symmetrized $n$-body
wave function where the bosonic or fermionic nature of any subset
of identical particles has to be taken into account in the
symmetrization (or anti-symmetrization) process 
in the standard manner \cite{gw}.

In this Letter we present explicitly the fully symmetrized wave function only 
for the case of $n$ identical charged bosons, as to our knowledge measurements
in high-energy physics attempting to reveal the strength of the 
multi-particle Bose-Einstein correlation
effects\cite{janus-na44,eggers} exist only for this special case. 

The fully symmetrized $n$-particle wave function has the form 
\begin{eqnarray}
\tilde \Psi^{(+){\cal S}}_{ {\rm {\bf {k}}}_{1} \cdots {\rm {\bf {k}}}_{n} }
({\rm {\bf {x}}}_{1}, \cdots ,{\rm {\bf {x}}}_{n}) 
& = & 
{1 \over \sqrt{n!} }
\sum_{\sigma^{(n)} }
\tilde \Psi^{(+)}_{ {\rm {\bf {k}}}_{1} \cdots {\rm {\bf {k}}}_{n} }
({\rm {\bf {x}}}_{\sigma_1}, \cdots ,{\rm {\bf {x}}}_{\sigma_n}) ,
\end{eqnarray}
where $\sigma^{(n)}$ stands for the set of permutations of $n$ different
indices, and $\sigma_i$ for the permuted value of the index $i$ in
one of the permutations belonging to the set $\sigma^{(n)}$.
Using the ansatz (\ref{aswn1}), the above equation simplifies as
\begin{eqnarray}
\tilde \Psi^{(+){\cal S}}_{ {\rm {\bf {k}}}_{1} \cdots {\rm {\bf {k}}}_{n} }
({\rm {\bf {x}}}_{1}, \cdots ,{\rm {\bf {x}}}_{n}) 
& = & 
{\sqrt{{\cal N}^{(n)}} \over \sqrt{ n!} }
\sum_{\sigma^{(n)} }
\prod_{i < j=1}^n {\psi}_{ {\rm {\bf {k}}}_{ij}}^{C(+)}
({\rm {\bf {r}}}_{\sigma_i \sigma_j}),
\label{wfsym}
\end{eqnarray}
which contains only the two-body relative Coulomb wave functions.

The physics of the above ansatz is very simple: if all $n$ final 
charges emerge
into the continuum and if all are well separated from the other tracks,
only the pairwise Coulomb relative wave functions play a r\^{o}le.
However, relative Coulomb wave functions have to be taken into
account for all possible particle pairs as the Coulomb interaction
is of long range. Graphically, if we represent the $n$ particles by
$n$ crosses, the relative Coulomb wave function 
between particle $i$ and $j$ can be represented by a line
connecting cross $i$ with cross $j$, and the full,
asymptotically correct $n$-particle Coulomb wave function
is represented by connecting each of the $n$ crosses with the 
$n-1$ others by forming a polygon with $n$ corners and
$n (n-1)/2 $ lines (diagonals and edges).

One can apply a simple approximation to 
(\ref{wfsym}) which preserves at least some features of 
the Coulomb distortion effects. 
It consists in neglecting, for any two-particle 
Coulomb wave function 
${\psi}_{ {\rm {\bf {k}}}_{ij}}^{C(+)}({\rm {\bf {r}}}_{ij})$, 
the hypergeometric function in the exact solution (\ref{2pwfn}) 
and retaining only the part 
$e^{ i {\rm {\bf {k}}}_{ij} {\rm {\bf {r}}}_{ij}} N_{ij}$. 
After evaluating the double sums over all permutations
of $\sigma_n$ in a product, one finds 
\begin{eqnarray}
\left|\tilde \Psi^{(+){\cal S}}_{ {\rm {\bf {k}}}_{1} 
\cdots {\rm {\bf {k}}}_{n} }
({\rm {\bf {x}}}_{1}, \cdots ,{\rm {\bf {x}}}_{n}) \right|^2
& = & {{\cal N}^{(n)} \over n!} 
\left( \prod_{i<j=1}^n G_{ij} \right) \,
\left| \sum_{\sigma^{(n)}} \prod_{i<j=1}^n
e^{ i {\rm {\bf {k}}}_{ij} {\rm {\bf {r}}}_{\sigma_i\sigma_j}} \right|^2 \, = 
\nonumber \\
\null & = & {{\cal N}^{(n)} \over {\cal N}_0 } 
\left( \prod_{i<j=1}^n G_{ij} \right) \,
\left|\Psi^{(0){\cal S}}_{ {\rm {\bf {k}}}_{1} \cdots 
{\rm {\bf {k}}}_{n} }
({\rm {\bf {x}}}_{1}, \cdots ,{\rm {\bf {x}}}_{n}) \right|^2.  
\label{river}
\end{eqnarray}
In the last line, the symmetrized $n$-particle wave function for
neutral particles $ \Psi^{(0){\cal S}}_{ {\rm {\bf {k}}}_{1} 
\cdots {\rm {\bf {k}}}_{n} }({\rm {\bf {x}}}_{1}, \cdots ,
{\rm {\bf {x}}}_{n})$ with its own normalization constant 
${\cal N}_0$ has been introduced. As usual,
\begin{equation}
G_{ij} := \left|N_{ij}\right|^2 = e^{- \pi \eta_{ij}}\, 
\left|\Gamma (1 + i \eta_{ij})\right|^2  \label{gij}
\end{equation}
is the Gamov penetration factor for the particle pair $(ij)$. 

It, thus, follows that the proper generalization of the Gamow penetration
factor for $n$ charged particles reads as
\begin{equation}
G_{1,\cdots, n} \, = \, \prod_{i < j = 1}^n G_{ij}.  \label{gn}
\end{equation}
This expression contains $n (n-1)/2$ factors, corresponding to all possible 
pairings $(ij)$. Moreover, it is self-consistent: if the momenta 
$k',k'',\cdots,k^{(l)}$ of $l$ particles approach infinity such 
that for no two momenta $k^{(i)}$ and $k^{(j)}$ the corresponding 
relative momentum remains finite, we have
\begin{equation}
\lim_{k' \to \infty} \cdots \lim_{k^{(l)} \to \infty } G_{1, \cdots, n} 
\, = \, 
G_{\alpha_1, \cdots , \alpha_{n-l}},
\end{equation}
where the remaining $n-l$ particles whose momenta remain finite are denoted 
symbolically by $\alpha_1, \cdots ,\alpha_{n-l}$.
Specifically,
\begin{equation}
\lim_{k_n \to \infty} G_{1, \cdots, n} 
\, = \, 
G_{1, \cdots , {(n-1)}}.
\end{equation}
An explicit check for $n=4$ shows that, indeed,
\begin{equation}
\lim_{k_4 \to \infty} G_{1,2,3,4} \, = \, 
\lim_{k_4 \to \infty} G_{12} G_{13} G_{14} G_{23} G_{24} G_{34} 
\, = \, G_{12} G_{13} G_{23}
\, = \, G_{1, 2, 3}.
\end{equation}
Hence, the generalization (\ref{gn}) of the Gamow correction 
factor to arbitrary 
values of $n$ is done in a self-consistent manner 
that satisfies its physically expected reduction property. 

This result was 
substantiated for the case of $n$ identical charged bosons.
In general, the final state of a high-energy heavy-ion reaction
contains many different kind of particles, with different charges
and quantum statistical properties. Nevertheless, the ansatz
given in Eq.\ (\ref{aswn1}) can be utilized for any values of the 
charges, and the result can be symmetrized for a generic mixture
of particles as prescribed in Ref.\cite{gw}. 

Let us add two comments: 

(i) As mentioned above, a wave function of the
type (\ref{aswn}) implies that the relative motion of each of the 
pairs of particles is independent of that of the other pairs, 
i.e., that no correlations between the motions of the particle 
pairs occur. In other words, the proposed form of the factorized 
$n$-particle Coulomb
wave function does not include genuine higher-order 
correlations, only those that can be built up from two-particle
Coulomb correlations. This is the same level of approximation
that is used to derive Eq.~(\ref{e:m2}), the generic form of the 
$n$-particle Bose-Einstein correlation functions. Results of
Ref.~\cite{cstjz,jzcst} suggest that Eq.~(\ref{e:m2}) is valid 
only if the density of bosons is below the limit of Bose-Einstein 
condensation.
 
(ii)  It should be kept in mind that 
the extrapolation out of the region $\Omega_{0}^{(n)}$ 
implied by (\ref{aswn}) is highly non-unique. 
Even for the 3-body Coulomb problem, various different 
wave functions which, of course, coincide asymptotically in $\Omega_{0}^{(3)}$  
with (\ref{aswn}) have been, and are still being, developed.

\section{Application to high-energy heavy-ion and particle collisions}

The correlation function measuring the enhanced probability for emission of 
$n$ identical Bose particles is given by Eq. (1).
 This correlation function
is usually, due to meager statistics, 
only measured as a function of the Lorentz 
invariant $Q_n$, defined by the relation
\begin{equation}
Q_n^2= \sum_{i<j = 1}^n q_{ij}^2
\end{equation}
where $q_{ij}={k}_i-{k}_j$, and where ${k}_i$ is the four-momentum 
of particle $i$.

\begin{figure}[htb]
\centering
\epsfig{file=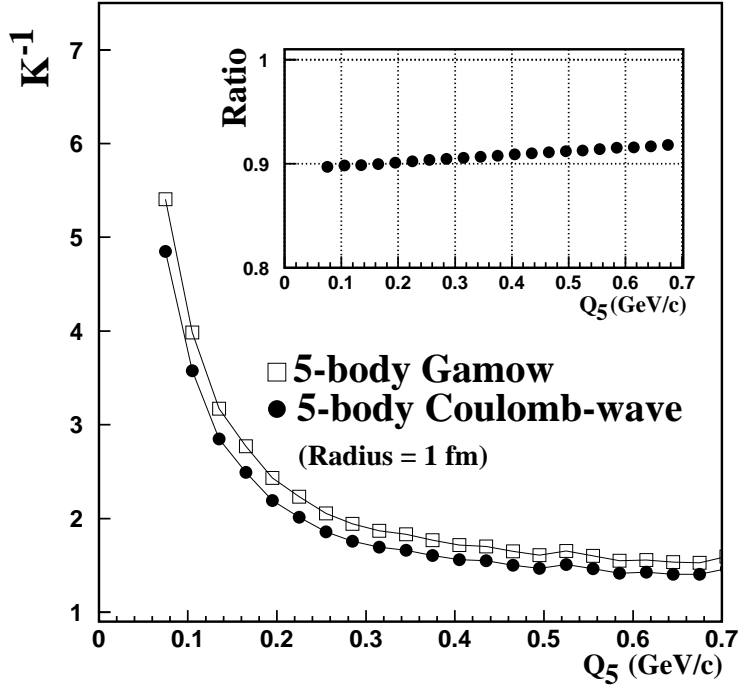,height=10.5cm}
\vspace{-0.5cm}
\caption{Filled circles stand for the Coulomb correction
	factor of 5-particle Bose- Einstein correlation functions for a
	source size of $R = 1 $ fm as obtained from the numerical integration of
	the 5-body Coulomb wave function, while the squares indicate
	the results of the less substantiated 5-body Gamow corrections;
	the inset shows the ratio of these two correction factors.
Lines are shown to guide the eye. }
\end{figure}

\begin{figure}[htb]
\centering
\epsfig{file=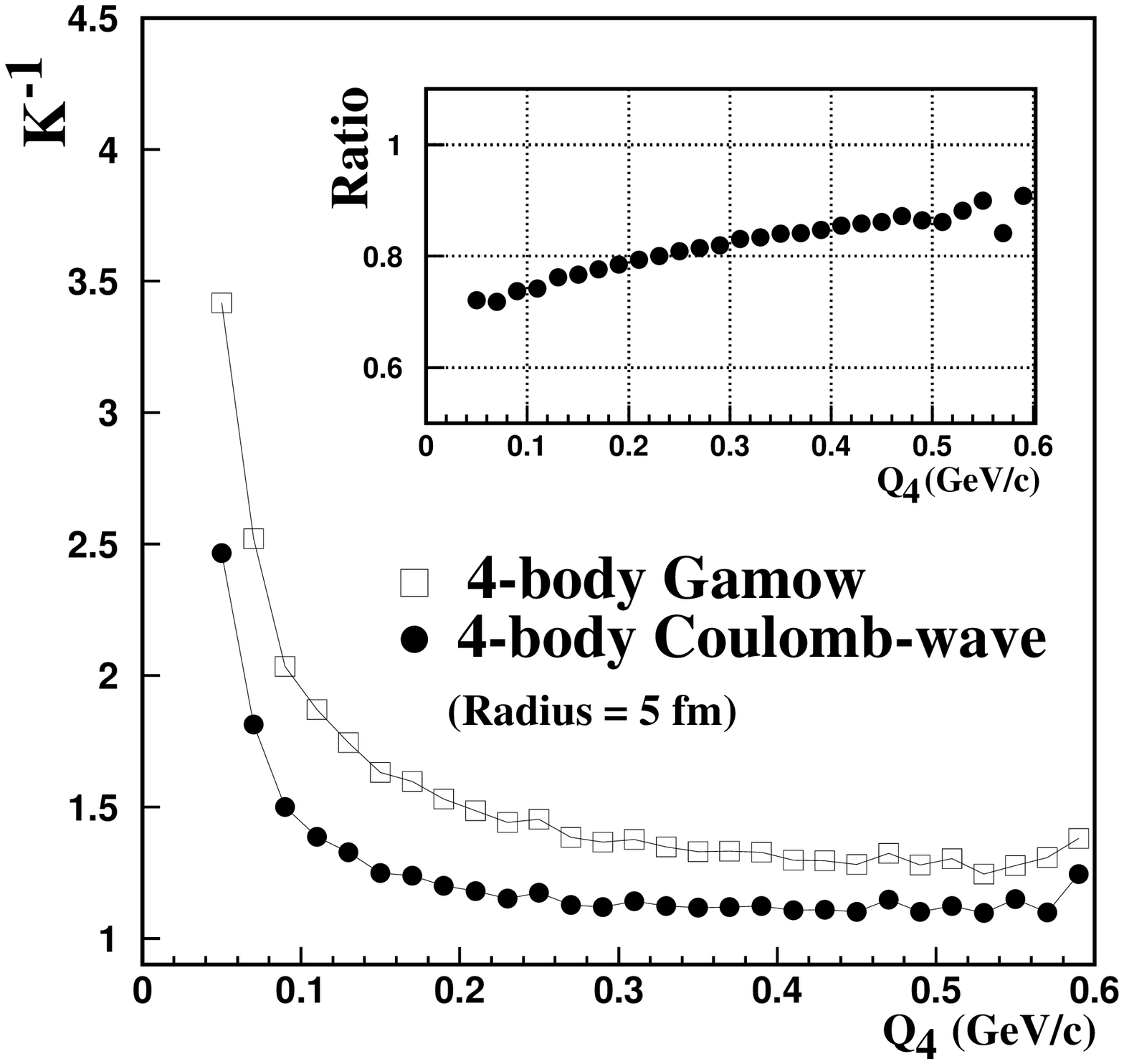,height=10.5cm}
\vspace{-1.0cm}
\caption{Same as Fig. 1 but for $n = 4$ particles and $R = 5$ fm}
\vspace{1.0cm}
\end{figure}

\begin{figure}[htb]
\centering
\epsfig{file=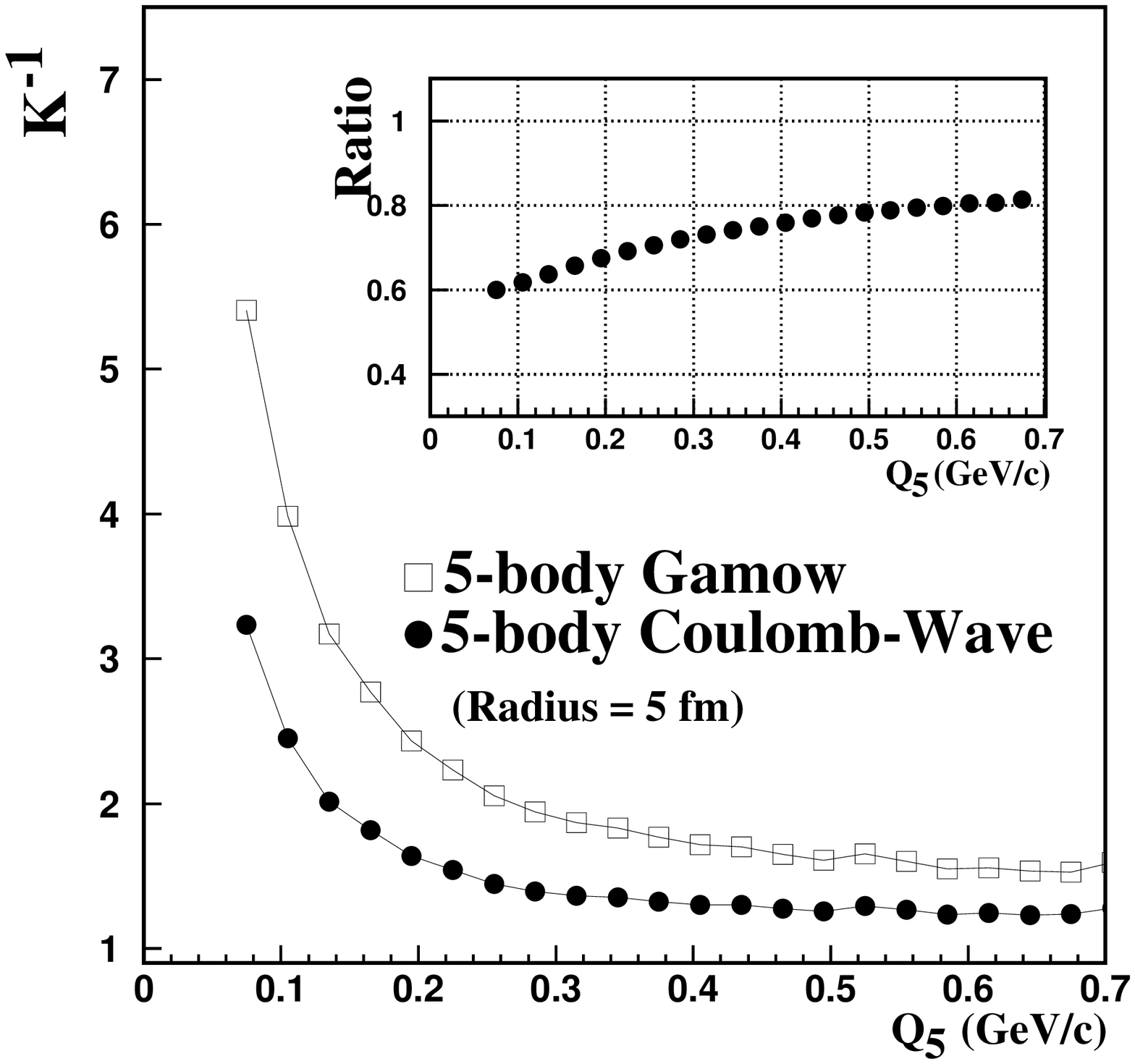,height=10.5cm}
\vspace{-1.0cm}
\caption{Same as Fig. 1 but for $n = 5$  and $R = 5$ fm}
\vspace{1.0cm}
\end{figure}

\begin{figure}[htb]
\centering
\epsfig{file=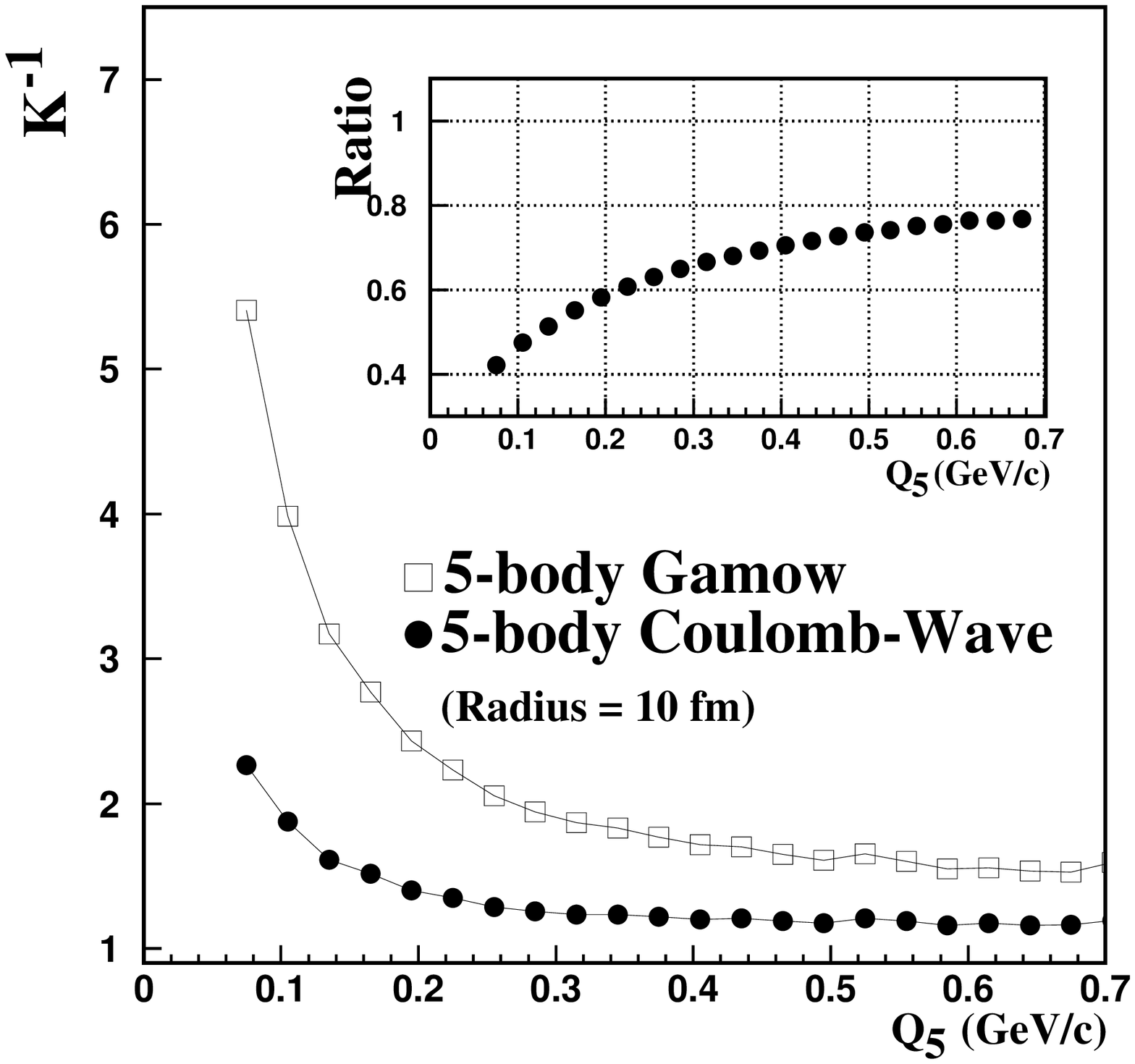,height=10.5cm}
\vspace{-1.0cm}
\caption{Same as Fig. 1 but for $n = 5$ and $R = 10$ fm}
\vspace{1.0cm}
\end{figure}

We can now calculate the Coulomb effects on 
the $n$-particle correlation function using 
\begin{equation}
K_{Coulomb}(Q_{n})= 
\frac{ \int \prod_{i = 1}^n d^{3}{\bf x_i} \rho ({\bf x_i})
\left|\tilde\Psi^{(+){\cal S}}_{ {\rm {\bf {k}}}_{1} 
	\cdots  {\rm {\bf {k}}}_{n} }
	({\rm {\bf {x}}}_{1},\cdots,
		{\rm {\bf {x}}}_{n})\right|^2}
	{ \int \prod_{i = 1}^n d^{3}{\bf x_i} \rho ({\bf x_i})
\left|\Psi^{(0){\cal S}}_{ {\rm {\bf {k}}}_{1} 
	\cdots {\rm {\bf {k}}}_{n} }
({\rm {\bf {x}}}_{1},\cdots, {\rm {\bf {x}}}_{n})\right|^2}
\label{Kcoul3},
\end{equation}
where $\rho({\bf x_i})$ is the density distribution 
of the source for particle $i$, 
taken as a Gaussian distribution of width R in 
all three spatial directions. 
This formulation makes it possible to 
extract information on  the source size $R$, 
and to compare this value with that extracted by means of
a generalized $n$-particle Gamow approximation 
through $K_{Coulomb}^{(G)}(Q_{n})= \prod_{i<j = 1}^n G_{ij}$. 
To this purpose we use the NA44 data sample of three pion 
events produced in S-Pb collisions at CERN \cite{janus-na44}.

We have calculated the Coulomb correction factor, 
i.e.\ $ K_{Coulomb}^{-1}(Q_{n})$~\cite{janus-na44}, 
for source radius values $R =$ 1, 5, and 10 fm, 
for $n = 2$, 3, 4, and 5 particle
correlations.
The radii were chosen to be in the range of interest 
for high-energy particle and high-energy heavy-ion physics. 
The results are compared  to the generalized Gamow approximation. 
We have checked that in the limit $R \rightarrow 0$ 
the $n$-particle Gamow approximation is indeed recovered numerically.

In case of a characteristic 1 fm effective source size,
typical for Bose-Einstein correlation functions in various elementary
particle reactions,
the difference between the $n$-particle Gamow and Coulomb wave function
corrections were smaller than 10 \% for $n = 4$ and 5 particles,
the $n$ = 5 case being shown in Fig. 1. However, for future
measurements of 5-particle Bose-Einstein correlations 
in particle physics that aim at a precision  better than 5 percent 
relative error, Coulomb wave function integration
will be a necessity.

For source sizes of 5 or 10 fm, that are the characteristic expectations
for Au + Au reactions at RHIC, the difference between the 
results of the Gamow and Coulomb wave function corrections increased
dramatically, see Figures 2 - 4. 
We find that for a source radius of 5 fm,
we need to take this detailed calculation into account 
already for the precise determination of the 
three-particle correlation function.
However, with increasing  number of particles,
the deviation between the $n$-particle Gamow and the 
$n$-particle Coulomb wave function integration method
increases drastically. For 5-particle Coulomb correction,
the better substantiated Coulomb wave function integration
method yields a factor of 2 deviation from the naive
generalized Gamow method.

At the end we note that systematic improvements of our treatment 
are possible by i) including also effects of strong interactions 
between the particles of each pair, ii) replacing the simple 
product of Gaussians by a more realistic model for the production 
of particle $n$-tuples, and iii) invoking improved $n$-body 
Coulomb wave functions that are correct in a larger region of 
$n$-body configuration space than $\Omega_{0}^{(n)}$. 

However, corrections i) and ii) are estimated to be small, as 
1) the final state Coulomb interaction dominates over the final 
state strong interaction due to its long range and the 
relatively large source size, and 2) the effective source of 
particles is known to be approximately Gaussian from detailed 
studies of two-pion correlations. Correction iii) is also expected 
to be small, as a clean measurement of particle $n$-tuples will 
likely require that these particles be in $\Omega_{0}^{(n)}$; however, 
one has to wait till 4-th and 5-th order correlations are measured 
in heavy-ion collisions in order to determine the more detailed 
experimental conditions.

\section{Summary and conclusions}

On the basis of an explicit, analytically given form of the 
$n$-body Coulomb wave function that is - at least asymptotically - 
correct in a large region of $n$-body configuration space, 
we have developed a new method to systematically correct for 
{\it explicit} many-body Coulomb effects which is 
applicable to data analysis in a broad range of measurements 
in high-energy particle and heavy-ion physics. A generalized
Gamow correction factor has been established as a limiting case of 
vanishing source sizes. 

Specifically, we have worked out 
our approach for 3, 4, and 5 identical charged particles 
and have tested it for Gaussian source sizes with $R = 1$, $5$,
and 10 fm.
We have numerically found that the 
generalized Gamow approximation is not reliable enough to 
determine the magnitude on the 5 \% level
of the five-body Coulomb correction factor 
if $R = 1$ fm, the characteristic lenght-scale of strong interactions
in high-energy particle physics.
The range of interest in high-energy
heavy-ion physics was probed in the 
$R = 5$ and 10 fm cases,  and systematic errors, as large as 100 \%,
were shown to be generated with the earlier Coulomb correction techniques for
the correlation function of 5 particles.

\section*{Acknowledgements} 
One of the authors (Cs. T.) would like to
thank Gy. Bencze and J. R\'evai for stimulating discussions. 
This research was partially supported by the Hungarian National Science
Foundation under grant OTKA T026435.

The results presented are the product of a workshop on the Coulomb 
three-body problem in high-energy physics held in Budapest September 
21-25, 1998. We are very grateful
for that intense and productive but at the same time relaxed week.
Support from the Swedish Research Council is acknowledged.


\begin{thebibliography}{99}
\bibitem{pnhalo} T. Cs\"org\H o, B. L\"orstad, J. Schmidt-S{\o}rensen, 
	A. Ster, Eu. Phys. J. C 9 (1999) 275. 
\bibitem{cstjz} T. Cs\"org\H o, J. Zim\'anyi,
	Phys. Rev. Lett. 80 (1998) 916.
\bibitem{Cramer} J. Cramer, K. Kadija, Phys. Rev. C 53 (1996) 908. 
\bibitem{nhalo} T. Cs\"org\H o, Phys. Lett. B 409 (1997) 11. 
\bibitem{biyajima} N. Suzuki, M. Biyajima,
	Phys. Rev. C 60 (1999) 34903. 
\bibitem{suzuki} M. Biyajima, A. Bartl, T. Mizoguchi, O. Terazawa, 
	N. Suzuki, Progr. Theor. Phys. 84 (1990) 931 ,
	Addendum - ibid. 88 (1992) 88. 
\bibitem{heinz} U. Heinz, Q. Zhang, Phys. Rev. C 56 (1997) 426.
\bibitem{axel}H. Heiselberg, A. P. Vischer, Phys. Rev. C 55 (1997) 874.
\bibitem{zhang} Q. H. Zhang, Phys. Rev. C 58 (1998) 18; 
		Nucl. Phys. A 634 (1998) 190.
\bibitem{eggers} H. C. Eggers, P. Lipa, B. Buschbeck, 
		Phys. Rev. Lett. 79 (1997) 197.
\bibitem{acls}  E. O. Alt, T. Cs\"org\H o, B. L\"orstad, 
		J. Schmidt-S{\o}rensen, Phys. Lett. B 458 (1999) 407. 
\bibitem{janus-na44} 
		J. Schmidt-S{\o}rensen et al, NA44 Collaboration,
		Nucl. Phys. A 638 (1998) 471 \\
		I. G. Bearden et al, NA44 Collaboration,
		Nucl. Phys. A 638 (1998) 103 ,\\
		H. B{\o}ggild et al, NA44 Collaboration,
		Phys. Lett. B 455 (1999) 77.
\bibitem{pratt} S. Pratt, Phys. Lett. B 301 (1993) 795.
\bibitem{gyu-pa} S. S. Padula, M. Gyulassy, Nucl. Phys. B 339 
		(1990) 378.
\bibitem{jzcst} J. Zim\'anyi, T. Cs\"org\H o, hep-ph/9705433, 
		Heavy Ion Phys. 9 (1999) 241. 
\bibitem{brood} T. Cs\"org\H o, J. Zim\'anyi, hep-ph/9811283,
	Proc. {\it Correlations and Fluctuations'98}: 
	8-th International Workshop on Multi-particle Production, 
        M\'atrah\'aza, Hungary, June 1998 
	(World Scientific, Singapore, 1999, ed. T. Cs\"org\H o, 
	S. Hegyi, R.  C. Hwa, G. Jancs\'o), p. 56.
\bibitem{a98b} E. O. Alt, Few-Body Syst. Suppl. 10 (1999) 65.
\bibitem{r72} P. J. Redmond, cited in 
		L. Rosenberg, Phys. Rev. D 8 (1972) 1833. 
\bibitem{am92} E. O. Alt, A. M. Mukhamedzhanov, 
		JETP Lett. 56 (1992) 435; 
		Phys. Rev. A 47 (1993) 2004.
\bibitem{dollard64} J. D. Dollard, J. Math. Phys. 5 (1964) 729; 
Rocky Mount. J. Math. 1 (1971) 5. 
\bibitem{Briggs90} J. S. Briggs, Phys. Rev. A 41 (1990) 539.
\bibitem{gw} M. L. Goldberger, K. M. Watson, 
		{\it Collision theory} (Wiley, New York, 1964). 
\bibitem{m77} S. P. Merkuriev, Theor. Math. Phys. 32 (1977) 680. 
\bibitem{bbk89} M. Brauner, J. S. Briggs, H. J. Klar,: J. Phys. B 22 (1989) 2265.
\end{thebibliography}
\end{document}